\pdfoutput=1
\documentclass{article}
\usepackage{preamble}
\title{Survey on Computational Applications of\\Tensor Network Simulations\footnote{Preprint version. Corresponding author: \texttt{marcos.diezgarcia@fujitsu.com}}}
\author[1]{Marcos Díez García}
\author[1]{Antonio M{\'a}rquez Romero}
\affil[1]{Fujitsu Research of Europe Ltd., Slough SL1 2BE, UK\\
{\texttt{marcos.diezgarcia@fujitsu.com}}\\
\texttt{antonio.marquezromero@fujitsu.com}}
\date{\today}
\begin{document}
\maketitle
\begin{abstract}
Tensor networks are a popular and computationally efficient approach to simulate general quantum systems on classical computers and, in a broader sense, a framework for dealing with high-dimensional numerical problems. This paper presents a broad literature review of state-of-the-art applications of tensor networks and related topics across many research domains including: machine learning, mathematical optimisation, materials science, quantum chemistry and quantum circuit simulation. This review aims to clarify which classes of relevant applications have been proposed for which class of tensor networks, and how these perform compared with other classical or quantum simulation methods. We intend this review to be a high-level tour on tensor network applications which is easy to read by non-experts, focusing on key results and limitations rather than low-level technical details of tensor networks.
\end{abstract}
\tableofcontents
\section{Introduction}\label{intro}
The roots of the computational applications we review can be traced back to a lecture given by Richard Feynman in the 1980s~\cite{Feynman1982}. It concerns the simulation {\linebreak} of subatomic-scale physical systems by means of computations based on Turing machines~\cite{Turing1936}, and the extent to which these machines can really simulate quantum physical systems was questioned. This motivated the proposal of quantum Turing machines by Deutsch~\cite{Deutsch1985}, from which quantum complexity theory~\cite{BernsteinVazirani1993} and quantum computing~\cite{NielsenChuang2010} developed into their present form.

Can classical computers \emph{efficiently} simulate quantum physical systems? This question is problematic because quantum mechanical theory represents quantum physical systems by vectors in a Hilbert space (assumed to be finite), which are called state vectors~\cite{Dirac1981,NielsenChuang2010}. Due to the quantum mechanical principle of superposition, state vectors are generally represented as linear combinations for a given basis in Hilbert space. This means two complex numbers are needed to represent state vectors of any two-state quantum physical system, or qubit, but also $2^n$ complex numbers are needed to represent state vectors of any composite {\linebreak} system with $n$ qubits. Just representing the states of quantum systems in a classical computer thus quickly becomes unfeasible for an increasing number of qubits, let alone simulate their behaviour over time.

A different but related question, which is a key open challenge in quantum computing, asks instead if classical Turing machines can efficiently simulate \emph{quantum circuits}~\cite{AaronsonChen2017,Daley2022}. Quantum circuits or, as originally called, quantum computational networks~\cite{Deutsch1989,NielsenChuang2010} are one of the de facto standard models of quantum computation for simulating quantum physical systems. Indeed, there exist certain quantum circuits that a classical Turing machine can efficiently simulate, for example: quantum circuits where all quantum gates are Clifford gates as shown by Gottesman in 1998~\cite{Gottesman1998} and quantum circuits based on so-called `matchgates' as shown by Valiant~\cite{Valiant2002} afterwards.

Another approach to efficiently simulate quantum circuits, not based on restricting the type of quantum gates, consists in representing quantum circuits as \emph{tensor networks} (TNs)~\cite{MarkovShi2008,Vidal2003}. TNs can operationally represent a quantum circuit by decomposing it into simpler circuit elements via tensors~\cite{Guo2021}. That is, quantum states, quantum gates and the operations between them are defined in terms of tensors and tensor algebra. A simple definition of tensor\footnote{Other mathematical notions and related historical remarks of tensors can be found in~\cite{Guo2021}.} is that of an ordered sequence of values indexed by zero, one or more indices; scalars, vectors and matrices are examples of a tensor. Thus a TN representing a given quantum circuit does not change \emph{what} the circuit computes but \emph{how} such computation is performed.

TNs made it possible to mathematically characterise other classes of quantum circuits that can be efficiently simulated on classical computers. One such class are quantum circuits with low entanglement~\cite{Vidal2003}; that is, where the number of entangled qubits grows at most polynomially with the total number of qubits. Another example is the class of quantum circuits where the number of gates grows polynomially, depth grows logarithmically and all qubit interactions are spatially localised~\cite{MarkovShi2008}. In fact, localised qubit interactions and low entanglement are properties of well-known antiferromagnetic material models in condensed matter physics~\cite{AKLT1987} that have been described via TNs~\cite{Orus2014}.

\subsection{Motivation and Contribution}\label{intro:contrib}
Many literature reviews have been published over the last decade discussing:
\begin{itemize}
    \item theory of TN methods~\cite{Bañuls2023,Bridgeman2017,OkunishiNishinoUeda2022,Orus2014,Orus2019}, of which the most widely used is the density-matrix renormalisation group algorithm~\cite{Schollwock2011,White1993};
    \item software for TN methods whether being standalone packages~\cite{Psarras2022} or part of larger software projects \cite{Sherrill2020,YoungSceseEbnenasir2023};
    \item partly theory and software of TNs \cite{Evenbly2022,GrasedyckKressnerTobler2013,KoldaBader2009};
    \item TN applications for one \emph{specific} research domain like numerical analysis {\linebreak} of continuous multivariate functions~\cite{Ripoll2021}, data analysis in machine {\linebreak} learning~\cite{Cichocki2017,Ji2023}, molecular orbitals in quantum chemistry \cite{BaiardiReiher2020,szalay2015tensor}, or {\linebreak} variational quantum algorithms in computational fluid dynamics~\cite{Jaksch2023}; and,
    \item simulation of quantum physical systems by means of classical or quantum computers though not specifically using TNs \cite{Daley2022,GeorgescuAshabNori2014,GyongyosiImre2019}.
\end{itemize}

Unlike the previous reviews, this paper presents a broad but not exhaustive review on state-of-the-art computational applications of TNs across many different research domains including: machine learning, mathematical optimisation, materials science, quantum chemistry and large-scale simulation of quantum {\linebreak} circuits. This choice of topics is motivated by the previous reviews and the overall aim of this paper. That is, to clarify what kind of applications have been proposed for which TNs, and if these TNs provide a performance advantage over other classical or quantum simulation methods. Our discussion throughout the review\footnote{We use the phrase `computationally intractable' to refer indifferently to a decision problem {\linebreak}or analogous optimisation problem for which no known algorithm (classical or quantum) solves it and whose computational time/space complexity is polynomially upper-bounded.} focuses on key results achieved by TN simulations and context related to the application itself, while technical details are kept to a bare minimum. We believe this will make it easier for readers who are not familiar with TNs but are interested in the applications.

This paper is organised as follows: Section~\ref{related} overviews other state-of-the-art {\linebreak} techniques, alternative to TNs, for classically simulating quantum circuits. {\linebreak} Section~\ref{overview} summarises basic notions of TNs used in subsequent sections. We then review applications based on well-known classes of TNs: image classification {\linebreak} (Section~\ref{ml}), mathematical optimisation (Section~\ref{opt}), material science and quantum chemistry (Section~\ref{materials}), and other emerging applications of TN simulations (Section~\ref{qc}). Finally, Section~\ref{discussion} concludes this paper with a general discussion of the applications reviewed including a summary table thereof.

\section{Related Work}\label{related}
This section contextualises our review by highlighting other popular approaches to simulate quantum circuits besides TNs.

Two general methodologies for classically simulating quantum circuits can be followed depending on two well known formulations of quantum mechanics: Schr{\"o}dinger's state-vector formulation~\cite{NielsenChuang2010} and Feynman's sum-over-paths (or path integral) formulation~\cite{Feynman1948}. Schr{\"o}dinger's formulation is the one followed by most quantum circuit simulators~\cite{YoungSceseEbnenasir2023}. A simulation here consists in state-vector transformations defined by the unitary operations of quantum gates~\cite{NielsenChuang2010}. By contrast, simulations under Feynman's formulation focus on computing single probability amplitudes (one for each possible measurement outcome) associated with the final quantum state of a given circuit. Although both methodologies {\linebreak} incur a computational time complexity that grows exponentially in the worst case, Feynman's requires a computational space complexity that only grows polynomially in the number of qubits and quantum gates~\cite{AaronsonChen2017}.

From those two methodologies, a range of specific and different techniques developed to classically simulate quantum circuits more efficiently, which we highlight as follows:
\begin{itemize}
    \item Exploiting \emph{massively parallel computing} made it possible in 2018 to sample {\linebreak} up to $2^{28}$ probability amplitudes from a state-vector simulation of quantum circuits with 64 qubits and a circuit depth of 22 \cite{Chen2018}, at a notable lower computational cost than before~\cite{DeRaedt2007}. Such number of qubits is {\linebreak} already beyond the 50-qubit scale that others in 2019~\cite{Wu2019} and 2022~\cite{Daley2022} thought to be the limit. To surpass it, however, required a combination of remarkable advances in parallel computer hardware using CPUs or GPUs, distributed or shared memory management as well as quantum circuit {\linebreak} partitioning techniques to distribute entries of state vectors and quantum gates' matrices across cluster nodes.
    \item Efficient \emph{data compression} techniques for floating-point data \cite{Wu2019} have been recently integrated into Intel-QS \cite{Guerreschi2020} distributed, full state-vector, classical simulator of quantum circuits. This enabled a leap in full state vector simulations from 45 qubits up to 61 qubits for Grover's quantum search algorithm and, at the same time, reduce memory usage from $32 \cdot 10^{18}$ bytes (without data compression) down to $768 \cdot 10^{12}$ bytes. Benchmarks up to 45 qubits across random circuit sampling, quantum approximate {\linebreak} optimisation algorithm and quantum Fourier transform, show that {\linebreak} memory usage can be reduced between 4.85 and 21.34 times the original thanks to such data compression while maintaining $97.6\%$ qubit simulation fidelity~\cite{Wu2019}.
    \item \emph{Decision diagrams} and \emph{TNs} are the most popular techniques to represent quantum circuits in a computationally more efficient way than using full state vectors~\cite{BurgholzerPloierWille2023}. Both use data structures allowing quantum circuits to be conveniently decomposed: TNs use tensors~\cite{Guo2021}, whereas decision diagrams use directed acyclic graphs similar to binary decision diagrams~\cite{Bryant1986}. But the simulation performance achieved by these data structures is rather dependent on the class of quantum circuits. For example, TNs are not {\linebreak} expected to perform well for deep and highly-entangled quantum {\linebreak} circuits~\cite{MarkovShi2008}. Decision diagrams are not expected to perform well if the original state vectors and quantum gates' matrices contain few redundant entries (e.g.~if most of all complex amplitudes are distinct)~\cite{MillerThornton2006,Niemann2016,ZulehnerWille2019}. The number of available quantum circuit simulators is significantly larger for TNs~\cite{Psarras2022}. Yet making an informed choice between them is challenging because the software is often redundantly developed and there is a lack of common development and documentation standards.
    \item More recently, several \emph{hybrid} techniques have been proposed to exploit the relative successes and computational performance trade-offs of the above techniques; for example: Schr{\"o}dinger-Feynman simulations via massively parallel computing~\cite{Markov2018} or decision diagrams~\cite{BurgholzerBauerWille2021}, tensor-based decision diagrams~\cite{Hong2022} and tensor-based circuit cutting~\cite{Guala2023}. Despite their promising results, these hybrid techniques are preliminary research, and their true performance advantages over the well established techniques above is yet to be clarified.
\end{itemize}

In contrast with classical simulation techniques, in 2016 the Institute of {\linebreak} Theoretical Physics in Zurich, Intel and Microsoft Research \cite{Haner2016} jointly proposed a classical \emph{emulator} of quantum circuits. Such emulator allows, in principle, to test and debug quantum circuits at a comparatively reduced computational cost than current classical simulators. This, however, involves a fundamental change of paradigm: the proposed classical emulator is required to compute the same output from a given quantum circuit as performed by a quantum computer {\linebreak} but, unlike classical simulators, it is not limited to do so by performing quantum {\linebreak} gate operations. Quantum gate logic can be replaced by faster, and functionally equivalent, classical subroutines to avoid overhead costs of simulating reversible gates as well as associated ancillary qubits. Benchmarks~\cite{Haner2016} for arithmetical {\linebreak} operations, quantum Fourier transform and quantum phase estimation show promising performance advantages of classical emulation over classical simulation of quantum circuits.

\section{Tensor Networks Overview}\label{overview}
This section defines the key notions of a tensor, tensor contraction and TN representations of state vectors, which are recalled in later sections.

An arbitrary tensor of complex numbers is an element $\mathbf{v}$ in a set $\mathbb{C}^{I_1 \times \cdots \times I_d}$, where $I_1, \ldots I_d$ are index sets such that $I_{j} = \{ 1, \ldots, m_j \}$ for all $j \in \{ 1, \ldots, d \}$ given any fixed natural numbers $m_j$ and $d$. Such tensor $\mathbf{v}$ is called an order-$d$ tensor because $d$ indices must be specified to retrieve a single complex number from $\mathbf{v}$. For example, an arbitrary matrix with two rows and three columns is specified as an order-$2$ tensor $\mathbf{v}$, where the rows are indexed by $I_1 = \{ 1, 2 \}$ with $m_1 = 2$ and the columns are indexed by $I_2 = \{ 1, 2, 3 \}$ with $m_2 = 3$. The element in the first row and third column is specified as $\mathbf{v}[1,3]$ with indices denoted in square brackets rather than subscripts. We use subscripts to label different tensors (e.g.~$\mathbf{v}_1$ and $\mathbf{v}_2$ are two different tensors).

Every state vector of a $n$-qubit composite system can be represented by a linear combination of basis state vectors in a product Hilbert space as
\begin{equation}\label{eq:1}
    \ket{\psi} = \sum_{\{i_1,\ldots, i_n \}} \mathbf{c}[i_1, \ldots, i_n] \left( \ket{i_1} \otimes \cdots \otimes \ket{i_n} \right) , \enspace \forall i_{k} \in I_k \textrm{ with } k = 1 \ldots n \enspace,
\end{equation}
where $\otimes$ denotes the Kronecker product, each $I_k$ is an index set, $\mathbf{c}$ is an order-$n$ tensor and each $\mathbf{c}[i_1, \ldots, i_n]$ is a complex number coefficient given by indices $i_1, \ldots, i_n$. To use computational basis states, one may choose $I_k = \{0,1\}$ so that $\ket{i_k}$ is either $\ket{0} = (1,0)$ or $\ket{1} = (0,1)$. For the case of a two-qubit system where $n = 2$, $$\ket{\psi} = \sum\nolimits_{\{i_1, i_2\}} \mathbf{c}[i_1,i_2] \ket{i_1, i_2}$$ expands into $$\ket{\psi} = \mathbf{c}[0,0] \ket{0,0} + \mathbf{c}[0,1] \ket{0,1} + \mathbf{c}[1,0] \ket{1,0} + \mathbf{c}[1,1] \ket{1,1}$$ using the shorthand notation $\ket{i_1, i_2} = \ket{i_1} \otimes \ket{i_2}$.

TN representations of $\ket{\psi}$ consist basically in expressing an order-$n$ tensor $\mathbf{c}$, see Equation~\eqref{eq:1}, in terms of lower order tensors which use fewer indices and thus are easier to handle.

\begin{figure}[H]
    \centering
    \includegraphics[scale=0.9]{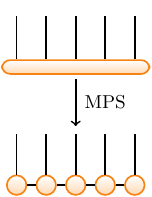}
    \caption{Diagram of a MPS representation (bottom) of an order-$5$ tensor (top). Every node corresponds to one tensor, every edge to one virtual index and every edge incident to only one node corresponds to one physical index.\label{fig:1}}
\end{figure}

\begin{figure}[H]
    \centering
    \includegraphics[scale=0.65]{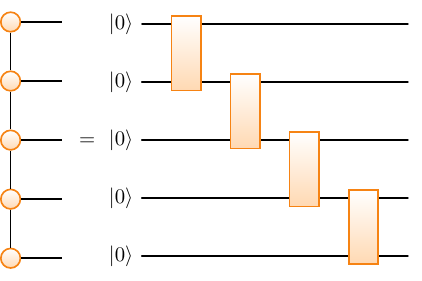}
    \caption{Quantum circuit diagram (right), consisting of two-qubit gates (`square boxes'), for a MPS of an order-$5$ tensor (left). \label{fig:2}}
\end{figure}

The simplest TN representation is called matrix product state (MPS) \cite{Orus2014,Vidal2003}. A general MPS representation of $\ket{\psi}$ is defined by expressing each coefficient $\mathbf{c}[i_1, \ldots, i_n]$ in Equation~\eqref{eq:1} as the following summation of tensor products
\begin{equation}\label{eq:2}
    \sum_{\{ \alpha_1, \ldots, \alpha_{n-1} \}} \mathbf{c}_{1}[i_1, \alpha_1] {\,} \mathbf{c}_{2}[i_2, \alpha_1, \alpha_2] \cdots \mathbf{c}_{n-1}[i_{n-1}, \alpha_{n-2}, \alpha_{n-1}] {\,} \mathbf{c}_{n}[i_n, \alpha_{n-1}] \enspace,
\end{equation}
where $\mathbf{c}_1$ and $\mathbf{c}_n$ are order-2 tensors (i.e.~matrices), and $\mathbf{c}_2$ through $\mathbf{c}_{n-1}$ are order-3 tensors. The indices $\alpha_1, \ldots, \alpha_{n-1}$ shared between tensors are called virtual or bond indices, which are mathematical artifacts of the representation and do not have a physical meaning per se. By analogy, indices $i_k$ are called physical indices because they relate to the physical degrees of freedom of a quantum state. Each $\alpha$ takes values in $\{1,\ldots, \chi \}$ where $\chi$ is a natural number called bond dimension. The bond dimension $\chi$ is normally considered as a parameter of the TN representation: increasing $\chi$ will increase the size of the corresponding tensor (i.e.~number of elements it contains). Equation~\eqref{eq:2} is a tensor contraction over such virtual indices. Note that multiplication of two matrices is a form of tensor contraction when any two given order-2 tensors, say $\mathbf{a}$ and $\mathbf{b}$, are contracted over a single shared index: $\mathbf{c}[u,v] = \sum_{w} \mathbf{a}[u,w] \mathbf{b}[w,v]$, where $\mathbf{c}[u,v]$ is the element at row $u$ and column $v$ of $\mathbf{c}$. Figure~\ref{fig:1} illustrates a MPS representation for a five-qubit system in diagrammatic form, and Figure~\ref{fig:2} illustrates a quantum circuit layout for such MPS representation by using two-qubit gates which act on consecutive pairs of qubits.

\begin{figure}[H]
    \begin{minipage}[b]{0.3\linewidth}
        \centering
        \includegraphics[scale=0.7]{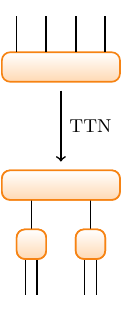}
         \subcaption{\label{fig:2:1}}
    \end{minipage}
    \begin{minipage}[b]{0.3\linewidth}
        \centering
        \includegraphics[scale=0.6]{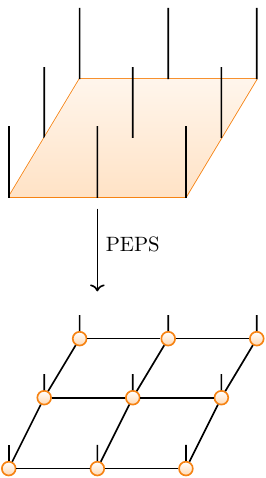}
        \subcaption{\label{fig:2:2}}
    \end{minipage}
    \begin{minipage}[b]{0.3\linewidth}
        \centering
        \includegraphics[scale=0.8]{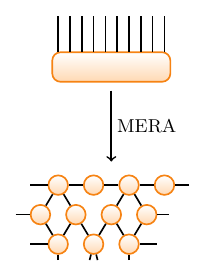}
        \subcaption{\label{fig:2:3}}
    \end{minipage}
    \caption{Diagrams of generalised tensor networks: tree tensor network (a), projected entangled-pair state (b), multi-scale entanglement renormalisation ansatz (c).\label{fig:3}}
\end{figure}

If one thinks of each tensor in a MPS as a `particle', then Figure~\ref{fig:1} clearly shows why MPS is a convenient representation of a quantum physical system where particles only interact with their nearest neighbours, as in the Affleck-Kennedy-Lieb-Tasaki model~\cite{AKLT1987,Orus2014}. Other quantum physical systems with more intricate interaction patterns can be represented and simulated by TNs which generalise MPS via higher order tensors. Well-known generalisations of MPS include: tree tensor network (TTN)~\cite{Shi2006,Tagliacozzo2009}, projected entangled-pair state (PEPS)~\cite{VerstraeteCirac2004}, a form of PEPS called isometric tensor network (isoTNS)~\cite{ZaletelPollmann2020} where tensors are similar to unitary matrices, and multi-scale entanglement renormalisation ansatz (MERA)~\cite{Orus2019}. Figure~\ref{fig:3} illustrates examples of these in diagrammatic form. Details about their formulations can be found in the aforementioned references.

The computational time and space complexities of TN representations generally depend on several factors including: the magnitude of the tensors' order {\linebreak} involved, the magnitude of the bond dimension, the order in which tensor products are performed during a contraction, and whether one truncates some of the tensors to approximately (rather than exactly) represent the original state vector $\ket{\psi}$. Finding an optimal way to contract a TN turns out to be {\linebreak} a NP-hard problem in general~\cite{GrayKourtis2021,MarkovShi2008}, as proved by reduction from the NP-completeness of the tree-width problem~\cite{Amir2010,Arnborg1987} or subset-sum problem~\cite{Chung1997}. {\linebreak} Nevertheless, it is sometimes possible to efficiently find `good' approximations of tensor decompositions by low-order tensors and contractions \cite{SilvaLim2008,Schuch2007}.

\section{Image Classification}\label{ml}
Many machine-learning applications have been developed between 2019 and 2023 for classifying two-dimensional digital images based on different TN representations:
\begin{itemize}
    \item Classification of handwritten digits taken from the modified National {\linebreak} Institute of Standards and Technology (MNIST) database \cite{Lecun1998}, which is a subset of the NIST Special Database 19 \cite{NIST2016}, by means of MPS~\cite{Wang2020} or TTN~\cite{ChenHao2023,LiuDing2019}.
    \item Classification of clothes taken from the Fashion-MNIST database~\cite{Xiao2017}, by means of MPS~\cite{Wang2020}, TTN~\cite{ChenHao2023} or PEPS~\cite{LiLai2023}.
    \item Classification of vehicles and animals taken from the ten-class Canadian {\linebreak} Institute for Advanced Research (CIFAR-10) database \cite{Krizhevsky2009}, by means of TTN~\cite{LiuDing2019}.
    \item Classification of COVID-19 pneumonia in X-ray chest images \cite{Chowdhury2020}, by means of PEPS~\cite{LiLai2023}.
    \item Classification of top quarks and discrimination of quantum chromodynamics background noise in calorimeter images, by means of MPS~\cite{ArazSpannowsky2021,ArazSpannowsky2022}, TTN~\cite{ArazSpannowsky2022} or MERA~\cite{ArazSpannowsky2022}.
\end{itemize}

The seminal work of Stoudenmire and Schwab in 2016~\cite{StoudenmireSchwab2016} introduced TNs for supervised machine learning. Their model consists in finding an optimal {\linebreak} weight tensor $\mathbf{w}$ for a real-valued decision function $f(x) = \mathbf{w} \cdot \Phi(x)$, which {\linebreak} classifies any input image $x = (x_1, \ldots, x_n)$ given by $n$ grey-scale pixels {\linebreak} $x_j \in (\mathbb{R} \cap [0,1])$. The weight tensor encodes the strength of correlations {\linebreak} between pixel values, and a MPS tensor network was proposed to represent it. The function $\Phi(x)$ is called feature map and transforms each pixel into a point with coordinates $\left( \cos \left( \frac{\pi}{2} x_j \right), {\,} \sin \left( \frac{\pi}{2} x_j \right) \right)$ in a unit circle, meaning that white pixels $x_j = 0$ map to the vector $(1,0)$ and black pixels $x_j = 1$ map to the vector $(0,1)$. These two vectors can be interpreted as qubits $|0 \rangle$ and $|1 \rangle$ respectively.

All the image classification applications reviewed in this section build upon Stoudenmire and Schwab's work, using different TN representations of the weight tensor $\mathbf{w}$. The main performance metric used to benchmark these TN models is the classification accuracy achieved on unseen data samples (i.e.~test accuracy). Other relevant aspects such as training time, tensor contraction time or memory usage are not benchmarked.

TNs have also been proposed for image generation, but this research is rather scarce and preliminary compared with image classification. Two examples are image generation for MNIST handwritten digits by means of TTN~\cite{Cheng2019} and generation of phase diagram images for a two-dimensional frustrated Heisenberg Hamiltonian by means of PEPS~\cite{Kottmann2021}. A general quantum machine learning model~\cite{Gao2018} was also proposed using a PEPS tensor network algorithm~\cite{SchwarzTemmeVerstraete2012}, which in theory can be applied to both classify and generate images.

\subsection{MNIST, Fashion-MNIST and CIFAR-10}\label{ml:mnist-fashion-cifar}
Novel TN representations may not always improve the test accuracy on certain benchmarks. For the MNIST dataset \cite{Lecun1998}, low-rank TTNs~\cite{ChenHao2023} obtain $98.3\%$ test accuracy (for bond dimension $\chi=8$ or higher), whereas hierarchical TTNs~\cite{LiuDing2019} obtain almost $95\%$ test accuracy (for bond dimension $\chi=10$). Yet both scores are lower than the $99\%$ test accuracy achieved already by Stoudenmire and Schwab~\cite{StoudenmireSchwab2016} for an MPS with a bond dimension of $\chi=120$.

In fact, achieving a test accuracy of $99\%$ on MNIST is not challenging, and it was already shown for a convolutional neural network called LeNet-5~\cite{Lecun1998} in 1998. Also, MNIST is not as computationally challenging for machine-learning methods as benchmarks like Fashion-MNIST~\cite{Xiao2017} and CIFAR-10~\cite{Krizhevsky2009} proposed more recently. On these, the test accuracy achieved by TN models is lower: multi-layered PEPS~\cite{LiLai2023} obtain $90.44\%$ on Fashion-MNIST (for bond dimension $\chi=5$); low-rank TTNs with tensor dropout~\cite{ChenHao2023} obtain $90.3\%$ on Fashion-MNIST (for bond dimension $\chi=16$); and, hierarchical TTNs~\cite{LiuDing2019} obtain $75\%$ on CIFAR-10 (for bond dimension $\chi=6$ or higher). For MPS with bond dimension $\chi=5$, an average classification accuracy of $92.2\%$ can be achieved on Fashion-MNIST~\cite{Wang2020} at a lower computational cost than said TTN and PEPS variants. However, this accuracy was measured via the AUROC\footnote{AUROC is a shorthand for `area under the receiver operating characteristic curve'~\cite{ChiccoJurman2023}.} metric, whose use is no longer recommended and can lead to overoptimistic results~\cite{ChiccoJurman2023}.

\subsection{COVID-19 Pneumonia}\label{ml:covid19}
Regarding the COVID-19 radiography dataset~\cite{Chowdhury2020}, multi-layered PEPS\footnote{Multi-layered PEPS~\cite{LiLai2023} are a generalised class of TN representations based on PEPS. It should not be confused with the use of the term `two-layer PEPS' in \cite{Pang2020} which refers to the contraction of two PEPS.} can achieve a test accuracy up to $91.63\%$ which is above the $87.08\%$ test accuracy by standard PEPS~\cite{LiLai2023}. This means using higher dimensional TNs, like multi-layered PEPS, can improve image classification accuracy over TNs with simpler structures. But the same authors~\cite{LiLai2023} show that image classification models based on multi-layered PEPS will also require more training parameters than those based on simpler TNs like PEPS, TTN or MPS. The authors report, for instance, that PEPS requires $1{\;}064{\;}964$ parameters for bond dimension $\chi=4$, whereas a two-layer PEPS requires $1{\;}394{\;}102$ parameters for bond dimension $\chi=3$ and $10{\;}750{\;}902$ parameters for bond dimension $\chi=5$. The convolutional neural network GoogleLeNet (Inception v1)~\cite{Szegedy2015} requires $6{\;}797{\;}700$ parameters\footnote{Authors~\cite{LiLai2023} incorrectly report $5{\;}604{\;}004$ as GoogleLetNet's total number of parameters.} in total and achieves a $92.75\%$ test accuracy on the COVID-19 radiography dataset~\cite{LiLai2023}. Therefore, GoogleLeNet can achieve higher classification accuracy using fewer training parameters than a multi-layered PEPS. Also, GoogleLetNet was released in 2015~\cite{Szegedy2015} and has been superseded by newer convolutional neural networks~\cite{ChenLeiyu2021}, so the performance trade-offs between multi-layered PEPS and simpler TNs or convolutional neural networks for classifying images are not clear yet.

\subsection{Top Quarks}\label{ml:quark}
In contrast with the purely classical models above, Araz and Spannowsky~\cite{ArazSpannowsky2022} {\linebreak} propose a quantum machine-learning model to classify the heaviest known {\linebreak} elementary particles, called top quarks, in images produced by calorimeters at {\linebreak} CERN's Large Hadron Collider. The ATLAS detector is used to generate such images, based on energy measurements from particles' collisions. In this model, input images are first encoded into an initial quantum state and then classified {\linebreak} by a quantum circuit, with gates arranged in a TN topology, according to a {\linebreak} decision function similar to Stoudenmire and Schwab's proposal~\cite{StoudenmireSchwab2016}. Araz and Spannowsky~\cite{ArazSpannowsky2022} showed that image classification by quantum circuits with MPS, TTN or MERA topologies involve notably fewer training parameters to achieve approximately the same or higher classification accuracy than a corresponding simulation on a classical computer (see Table~\ref{table:1}). However, the accuracy was measured via the not so reliable AUROC metric~\cite{ChiccoJurman2023}, and therefore the validity of Araz and Spannowsky's benchmark results needs clarification.
\begin{table}[H]
    \small
    \centering
    \begin{tabular}{@{}c c c c@{}}
        \toprule
        Simulation
        & Tensor Network
        & \# Training Parameters & Best Accuracy (\%)
        \\ \toprule
        \multirow[c]{3}{*}{Classical}
        & MPS
        & 2150
        & 89.4 \\
        & TTN
        & 14800
        & 89.6 \\
        & MERA
        & 18200
        & 90.1
        \\ \midrule
        \multirow[c]{3}{*}{Quantum Circuit}
        & MPS
        & 9
        & 88.6 \\
        & TTN
        & 9
        & 89.3 \\
        & MERA
        & 17
        & 91.4
        \\ \bottomrule
    \end{tabular}
    \caption{Classification accuracy by tensor networks for top quark discrimination from background radiation noise in two-dimensional calorimeter images. \label{table:1}}
\end{table}

One challenge faced by machine-learning models, particularly those based on gradient-descent methods, are barren plateaus or flat regions associated with the loss function used for training the model (see Section~\ref{opt:vqa-plateaus}). However, Araz and Spannowsky~\cite{ArazSpannowsky2022} also argue that their classification model is unlikely to suffer from barren plateaus if simulated on a quantum circuit whether using MPS, TTN or MERA topologies. To show this, the authors analyse the eigenvalue distribution of the empirical Fisher information matrix~\cite{Abbas2021} for the quantum TNs and their corresponding simulation on a classical computer. Such eigenvalue distribution can be used as an indirect measure of flatness in the loss function~\cite{Abbas2021}: a model suffering from barren plateaus will have an increasing number of said eigenvalues around zero as the number of qubits of the model increases.

\section{Optimisation and Local Hamiltonians}\label{opt}
A class of optimisation problems that is central to quantum computing consists in finding a minimum eigenvalue $\lambda$ and associated eigenvector $\vert  \psi \rangle$ of a given Hamiltonian operator $H: \mathcal{H}^{\otimes n} \to \mathcal{H}^{\otimes n}$, on a Hilbert space $\mathcal{H}$ of $n$-qubit states, so that $H \vert  \psi \rangle = \lambda \vert  \psi \rangle$ holds \cite{Bharti2022,KempeKitaevRegev2006,NielsenChuang2010}. $H$ represents the energy function of a physical system, mapping a given ground state $\vert  \psi \rangle$ to a corresponding ground-state energy value $\lambda$. Usually, the Hamiltonian can be expressed as a sum of sub-functions $H = \sum_{j=1}^{r} {\,} H_j$, where each sub-function $H_j$ is locally defined on $k$ qubits at most (given a fixed $k \leq n$) and the number of terms $r$ is polynomial in $n$. In other words, one often can expand Hamiltonians as a finite series without exponentially many terms and limit the maximum number of interacting qubits to $k$. This refers to the class of $k$-local Hamiltonians~\cite{KempeKitaevRegev2006}, which is a generalisation of maximum $k$-satisfiability problems for the quantum complexity class of QMA-complete problems (analogous to the NP-complete complexity class).

The subclass of 2-local Hamiltonians is noteworthy for two reasons. First, this subclass suffices to prove that any adiabatic quantum computation, {\linebreak} performed by quantum annealers~\cite{AlbashLidar2018} for instance, can be efficiently simulated by {\linebreak} the quantum circuit model and vice versa~\cite{KempeKitaevRegev2006}. This equivalence suggests that, for all $k$-local Hamiltonians with $k \geq 2$, TN simulations of a quantum circuit {\linebreak} can provably be performed by an adiabatic quantum computer (though this review focuses on quantum circuits). Second, 2-local Hamiltonians include {\linebreak} popular Hamiltonian models such as the Lenz-Ising model of spin glasses based on Sherrington and Kirkpatrick's work~\cite{SherringtonKirkpatrick1975}. This model has applications in physics, chemistry, biology and combinatorial optimisation~\cite{Stadler1995}. In fact, many constrained optimisation problems can be reformulated as a spin-glass model or a quadratic unconstrained binary optimisation (QUBO) problem~\cite{GloverKochenbergerDu2019,Lucas2014,Stadler1995}.

In the following, we cover TN applications in QUBO and related methods for dynamic portfolio problems in finance~\cite{Mugel2022}, a variational quantum algorithm for QUBO problem solving~\cite{Perelshtein2023} and analysis of barren plateaus in variational quantum optimisation~\cite{MiaoBarthel2024}. Besides these, one can find spin-glass models for the so-called dose optimisation in cancer radiotherapy, which can be solved via a TTN algorithm~\cite{Cavinato2021}. This is preliminary research, and we exclude it for the lack of clear performance advantages over state-of-the-art algorithms and comparison with alternative TN representations.

\subsection{Variational Quantum Optimisation and Barren Plateaus}\label{opt:vqa-plateaus}
Variational quantum algorithms (VQAs) are hybrid quantum-classical methods which iteratively try to approximate an optimal solution to QUBO problems, spin-glass models or $k$-local Hamiltonian problems more generally ~\cite{Bharti2022,Tilly2021}. In essence, VQAs use a parametrised quantum circuit to first generate an initial quantum state and compute its energy for a given Hamiltonian as an expectation value. This estimated value is used as an upper-bound of the ground-state energy or optimal solution. Based on this guess, a separate search algorithm run on a classical computer will heuristically update the circuit parameter values to generate an improved guess in the next VQA iteration until some convergence criteria is satisfied.

Unfortunately, convergence of VQAs towards an optimal solution can fail because of barren plateaus \cite{McClean2018} among other major issues such as quantum noise, parameter initialisation and quantum state initialisation~\cite{Bharti2022,Scriva2024,Tilly2021}. Put simply, a barren plateau occurs when the expectation value of the energy or cost function resembles a flat surface, which leaves the VQA with no useful heuristic information (e.g.~gradients) to update the circuit parameters and possibly leads to random search behaviour~\cite{McClean2018}.

Nevertheless, recent research~\cite{Cerezo2021} strongly suggests that barren plateaus can {\linebreak} be avoided for certain VQAs if:
\begin{enumerate*}[(a)]
    \item the circuit depth is restricted to grow no faster than logarithmically in the number of qubits (i.e.~shallow circuits); and,
    \item the maximum number of qubit interactions at any one time is fixed to a finite and preferably `small' number (e.g.~nearest neighbours), thus restricting {\linebreak} the class of cost functions and associated optimisation problems that the VQA can possibly solve.
\end{enumerate*}

These conditions were assumed by separate research to prove the absence of barren plateaus in TN representations of quantum circuits with MPS~\cite{Leone2022}, TTN and MERA topologies~\cite{MiaoBarthel2024}. Moreover, shallow quantum circuits with local interactions can be efficiently simulated on classical computers by means of TN contractions~\cite{MarkovShi2008,Vidal2003}. In short, the topology of TNs and the contraction computational complexity can provide valuable analytic insight about the presence of barren plateaus in VQAs.

There exist examples of VQAs like `QuEnc'~\cite{Perelshtein2023} that are specifically {\linebreak} designed to solve binary optimisation problems with simple linear equality {\linebreak} constraints by reformulating them as QUBO problems. Using a MPS representation, the authors show that the running time of classically simulating a (shallow) five-layer QuEnc circuit scales linearly as the number of qubits increases, taking no more than one second on a regular laptop for 300 qubits~\cite{Perelshtein2023}. In terms of QUBO problem solving, authors also benchmarked QuEnc using a real quantum {\linebreak} computer (specifically, IBM's five-qubit \texttt{ibmq{\_}manila}) and compared it against simulated annealing (SA) and Goemans-Williamson's (GW) algorithm. Based on randomised 256-nodes graph instances of the maximum cut problem, QuEnc's solution quality improves as its circuit depth is increased from five to 20; {\linebreak} however, even with a 20-layer QuEnc circuit, the solution quality is still worse than both SA and GW.

\subsection{Dynamic Portfolio Optimisation}\label{opt:portfolio}
The mean-variance model introduced by Markowitz~\cite{Markowitz1952} is the basis of current {\linebreak} quantitative approaches to portfolio selection used in finance. It is formalised as a constrained optimisation problem where the goal is to find a portfolio, that is a vector of proportions of a given capital for investment across assets, {\linebreak} which maximises the expected return on the investment while minimising {\linebreak} financial risk. Optimal solutions in Markowitz's model can be found efficiently with classical solvers, but newer realistic models based on discrete formulations {\linebreak} with additional constraints turn portfolio optimisation into a mixed-integer {\linebreak} programming problem that is computationally intractable~\cite{Bienstock1996,JinQuAtkin2016}.

One example are dynamic portfolio optimisation models, where portfolios are generalised from single-period to multi-period investments over a series of consecutive trading days. Researchers from Multiverse Computing~\cite{Mugel2022} benchmarked a MPS-based algorithm as well as state-of-the-art quantum and classical {\linebreak} solvers against dynamic portfolio problems formulated as QUBO with up to 1272 variables and all-to-all interaction pattern. The benchmark results show that the MPS algorithm achieved the best Sharpe ratios. Sharpe ratios measure {\linebreak} solution quality as the proportion of expected return per unit of risk. The MPS algorithm also outperformed, regarding problem size scalability, a classical solver provided by Python's GEKKO library and two VQAs (implemented by the authors using IBM's quantum platform and Xanadu's PennyLane library). However, the MPS algorithm performed worse than GEKKO's classical solver in terms of solution quality measured via total profit (i.e.~returns minus transaction costs) and worse than D-Wave's 2000Q quantum annealer in terms of running time. Overall, the best trade-off between Sharpe ratios and speed for solving dynamic portfolio optimisation problems is attained by the MPS algorithm and D-Wave's 2000Q quantum annealer.

\section{Materials Science and Quantum Chemistry}\label{materials}
This section covers several applications proposed between 2013 and 2023 for analysis and discovery of materials, all of which are based on computing ground states of local Hamiltonians via classical TN simulations.

\subsection{Artificial Graphene}\label{materials:graphene}
Graphene is a prime example of nanomaterial that is made of a single layer of carbon atoms forming a two-dimensional hexagonal structure. Graphene has {\linebreak} many applications in energy storage, steel coating and biomedical sensors~\cite{Choi2010} despite posing risks for biological systems \cite{Seabra2014}. The so-called artificial graphenes are materials with graphene-like properties that can be manufactured using {\linebreak} alternative substrates such as alluminium gallium arsenide \cite{Gibertini2009}.

A quantum circuit was recently proposed to find the ground state of artificial graphene~\cite{Obiol2022} based on a 2-local Hamiltonian model proposed by Hubbard~\cite{Hubbard1963}. The proposed circuit's depth grows linearly with the number of qubits, which suggests the circuit is not unreasonably deep even though TNs are more efficient on shallow circuits where depth grows logarithmically~\cite{MarkovShi2008,MiaoBarthel2024,Perelshtein2023}. Finding the exact ground state by diagonalising the Hamiltonian exceeds the memory limitations of the MareNostrum 4 supercomputer for circuits beyond 20 qubits or graphene lattices with more than two hexagons~\cite{Obiol2022}. However, the same authors also show that approximating the ground state via full state vector simulation is possible up to 32 qubits and, if using a MPS representation of the circuit, up to 36 qubits with $1\%$ accuracy error relative to the true ground state. This result demonstrates that high-accuracy classical simulations of Hubbard's model are possible beyond the 24 qubit limit reached in past experiments with VQAs using no TN representation~\cite{Cade2020}.

\subsection{Hydrogen Chains, Ethane and Atazanavir}\label{materials:hydrogen-ethane}
China's fastest supercomputer, Sunway TaihuLight (SW26010 Pro), has been recently used to classically simulate a MPS-based VQA~\cite{shang2023towards} for finding ground states of the following molecules: hydrogen, ethane, hydrogen chain of 500 atoms, and atazanavir which is a prescription medicine to treat the human immunodeficiency virus. According to the authors~\cite{shang2023towards}, these are the largest quantum-circuit simulations reported to date for a quantum chemistry problem in terms of the total number of qubits ($n$) or CNOT gates ($n_{\textrm{CNOT}}$) involved:
\begin{itemize}
    \item hydrogen \ch{H2}, $n = 92$, $n_{\textrm{CNOT}} = 1.4 \cdot 10^{5}$;
    \item ethane \ch{C2 H6}, $n = 32$, $n_{\textrm{CNOT}} = 4.4 \cdot 10^{5}$;
    \item hydrogen chain $(\ch{H2})_{250}$, $n = 1000$, $n_{\textrm{CNOT}} = 10^{6}$; and,
    \item atazanavir \ch{C38 H52 N6 O7}, $n = 16$, $n_{\textrm{CNOT}} = 1.8 \cdot 10^{6}$.
\end{itemize}

The benchmarks for hydrogen and hydrogen chain molecules~\cite{shang2023towards} show that MPS-VQA achieves enough chemical accuracy to match exact reference values of ground energies (i.e.~full configuration interaction) obtained via Python's PySCF library. However, such level of chemical accuracy is not reported for neither ethane nor atazanavir molecules; in fact, the same authors suggest that improved accuracies can be achieved by using other VQA designs. Moreover, a single iteration of their proposed MPS-VQA takes more than 30 minutes to complete, using 512 cores of Sunway's supercomputer, for a hydrogen chain of 500 atoms. Taking more than 30 minutes for only one MPS-VQA iteration is arguably a long running time, which agrees with the fact that TNs like MPS are not adequate for such deep VQA circuits (see Section~\ref{opt:vqa-plateaus}). But, more importantly, the benchmarks do not show what performance advantages does the proposed MPS-VQA provide over other state-of-the-art methods (whether classical, quantum, based on TNs or not). Full state-vector simulations of many useful quantum circuits, including VQAs, with 45 qubits and more have been successfully demonstrated in 2019~\cite{Wu2019}.

\subsection{Tree-shaped Molecules}\label{materials:treeshape}
MPS is the most common representation used in TN algorithms to find ground states of quantum chemistry Hamiltonians~\cite{BaiardiReiher2020,szalay2015tensor}. However, the electronic interaction pattern in certain tree-shaped molecules is not accurately described by the linear structure that is characteristic of MPS. Performance advantages of TTNs, in terms of lower running times or lower energy estimation errors during simulation, have been demonstrated using toy examples for crystalline salts like lithium fluoride~\cite{Murg2015} and Cayley trees formed by hydrogen atoms~\cite{NakataniChan2013} as {\linebreak} well as more realistic examples using nitrogen dimers and stilbenoid {\linebreak} dendrimers (naturally occurring in plants) with up to 110 electrons and 110 active orbitals~\cite{NakataniChan2013}.

A decade later, research still continues to characterise the classes of quantum circuits for which TTN can outperform MPS~\cite{Seitz2023}. Using a single CPU (AMD Ryzen 7 3700) and 32 GB of RAM, it has been shown experimentally~\cite{Seitz2023} up to 37 qubits that TTN scales exponentially better than MPS in terms of wall-clock time and bond dimension provided that: the circuits exactly match a well defined tree layout, are shallow and have limited entanglement. But this result was obtained on artificial and carefully chosen quantum circuits that were an ideal fit for TTN. It is far from clear if and how such TTN performance advantage can be extrapolated to other problems involving approximately the same number of qubits, like artificial graphene with 36 qubits (Section~\ref{materials:graphene}) or ethane with 32 qubits (Section~\ref{materials:hydrogen-ethane}), for which top-class supercomputers were needed.

\subsection{Discovery of Physical Phases}\label{materials:phases}
Finding the ground state of a physical system can be a challenging task not just because it entails solving a local Hamiltonian problem~\cite{Bharti2022,KempeKitaevRegev2006}, which is {\linebreak} computationally intractable in general, but also because the ground state itself can vary depending on whether the physical system at hand undergoes sudden phase transitions according to changes in pressure, temperature or a magnetic field force for example. Thus analysing such phase transitions is a fundamental {\linebreak} part of research in quantum computing and quantum physics at large. In fact, the existence of exotic phases like topological quantum phases~\cite{Wen2017} lays a theoretical foundation to build universal quantum computers which are intrinsically and fully fault-tolerant at hardware level~\cite{Field2018}.

Quantum Monte Carlo (QMC) algorithms have been used to find phase transitions of Hamiltonian models but certain shortcomings of QMC recently motivated the use of TNs, such as PEPS for the Shastry-Sutherland model~\cite{CorbozMila2014} and two-dimensional isometric TNs (similar to PEPS) for the transverse-field Ising model~\cite{KadowPollmanKnap2023}. These TN applications are part of fundamental research to develop new technologies. For instance, the Shastry-Sutherland model~\cite{ShastrySutherland1981}, with spins arranged on a two-dimensional lattice with next-nearest neighbour interactions, is the only known model for which it is possible to find the exact ground states of an alkaline earth oxide material called strontium copper borate \ch{Sr Cu2 (BO3)2}. This material is relevant because it is thought to be a Mott-Hubbard insulator that can exhibit superconductivity~\cite{Kageyama1999,ShastryKumar2002}, and previously unknown phases of \ch{Sr Cu2 (BO3)2} have been discovered using PEPS \cite{CorbozMila2014,Shi2022}.

\section{Other Trends in Tensor Network Simulations}\label{qc}
This section covers emerging TN applications, proposed between 2017 and 2024, for other selected topics: computational fluid dynamics (Section~\ref{qc:cfd}), quantum advantage experiments (Section~\ref{qc:advantage}) and quantum error correction (Section~\ref{qc:qe}).

\subsection{Computational Fluid Dynamics}\label{qc:cfd}
The Navier-Stokes equations are non-linear partial differential equations that have been traditionally used to model the time-dependent behaviour of fluids.{\linebreak} Except when simplifying assumptions are made, obtaining solutions to such equations by traditional methods like direct numerical simulation (DNS) is {\linebreak} computationally inefficient for classical computers. This is particularly evident if one considers realistic turbulent flows with complex geometries characterised by high Reynolds numbers~\cite{MoinMahesh1998}.

Kiffner and Jaksch~\cite{KiffnerJaksch2023} propose an alternative and general DNS method based on a MPS representation that encodes the velocities describing flow {\linebreak} dynamics. They illustrate it for a simple toy model solving the incompressible {\linebreak} Navier-Stokes equations, called lid-driven cavity model, where the flow is {\linebreak} confined to a (discrete) square lattice in two spatial dimensions and the flow density does not change over time. To justify the computational efficiency of this approach, the authors empirically show that
\begin{enumerate*}[(a)]
    \item the number of parameter {\linebreak} variables describing the flow grows proportionally to the bond dimension of MPS, and
    \item the bond dimension grows logarithmically with simulation time.
\end{enumerate*}
This leads to faster runtimes compared with DNS for high Reynolds numbers (Re), peaking at 17 times faster for $\textrm{Re} = 60.5 \cdot 10^{3}$. All benchmarks comparing DNS and their MPS-based method are implemented using MATLAB and run on a single CPU node (Intel Xeon Platinum 8268) of Oxford's Advanced Research Computing facility. However, authors warn that the performance advantage of MPS may degrade if simulation time or bond dimension increases significantly.

\subsection{Quantum Advantage Experiments}\label{qc:advantage}
Quantum computers are expected to perform tasks which are computationally intractable for classical computers, even though it remains unclear which task is most appropriate to benchmark such quantum advantage as well as what quantum computer implementation can achieve it in practice and at what cost. Over two decades of research advances towards fault-tolerant quantum computation~\cite{Shor1996} elapsed, yet all current physical implementations of quantum {\linebreak} computers perform noisy and error-prone quantum computations~\cite{Bharti2022,Daley2022}. These are often called noisy intermediate-scale quantum (NISQ) computers without a commonly agreed and exact definition of how noisy or large. 

One benchmark task for demonstrating a quantum advantage is sampling fixed-length bitstrings from the output of a pseudo-random quantum circuit, as popularised by an experiment in 2019 on Google’s Sycamore superconducting {\linebreak} quantum processor with 54 qubits arranged on a rectangular lattice with nearest-neighbour interactions~\cite{Arute2019}. This task is regarded as computationally intractable for classical computers mainly due to the highly-entangled quantum states {\linebreak} generated by such random quantum circuits. Google researchers~\cite{Arute2019} estimated that it would take $10{\ }000$ years for state-of-the-art supercomputers to compute one million samples from a random quantum circuit with 53 qubits and a depth of 20. However, recent advances on massively parallel and efficient PEPS-based simulators~\cite{guo2019general,Liu2021} have shown that the same task can be classically simulated within 304 seconds on the Sunway TaihuLight (SW26010 Pro) supercomputer. The largest scale achievable by these PEPS simulations are random quantum circuits with a $10 \times 10$ grid of qubits and depth of $42$, well beyond Google's Sycamore experiment.

Another attempt to empirically show a quantum advantage was conducted in 2023 for computing expectation energy values of a two-dimensional transverse-field Ising Hamiltonian model~\cite{Kim2023}. Here IBM’s Eagle superconducting quantum processor with 127 qubits is benchmarked against MPS and two-dimensional isoTNS classical simulators run on a single 64-core processor and 128 GB of memory. The transverse-field Ising model is chosen because it matches the IBM Eagle processor's topology. The authors~\cite{Kim2023} argue running quantum circuits of that many qubits, with up to 60 layers of two-qubit gates and $2880$ CNOT gates, is out of reach for classical simulators. Once again, however, a remarkable {\linebreak} follow-up work~\cite{Patra2024} showed for the same Ising model that a PEPS-based {\linebreak} classical simulator not only can efficiently and accurately simulate IBM's {\linebreak} Eagle processor but also IBM's Osprey and Condor newer quantum processors of 433 and 1121 qubits respectively. Similar experiments demonstrating efficient simulations of TNs for the same transverse-field Ising model on 127 qubits have been conducted independently by other researchers~\cite{Begusic2024,Liao2023,Tindall2024}. An exception is D-Wave's recent experiment for a transverse-field Ising model~\cite{King2024}, claiming {\linebreak} that no classical simulation of MPS nor PEPS on Summit and Frontier {\linebreak} supercomputers can match in practice the same accuracy as D-Wave's quantum{\linebreak} annealers Advantage and an Advantage2 prototype (with 5627 and 1222 qubits respectively).

The above unprecedented results provide evidence for the utility of PEPS-based simulators and refine the current benchmark baselines which future {\linebreak} experiments will have to surpass to show quantum advantage.

\subsection{Quantum Error Correction}\label{qc:qe}
Performing quantum computations at arbitrarily large scales beyond what can be efficiently simulated by classical computers is key to show a practical quantum {\linebreak} advantage. Demonstrating this, however, poses major challenges due to the presence of quantum noise which can destroy information encoded in quantum states and thus corrupt the result of quantum computations. Quantum error {\linebreak} correction (QEC) methods enable fault-tolerant quantum computing at the {\linebreak} expense of using many redundant physical qubits to implement a single, error-corrected, logical qubit \cite{Bharti2022,Daley2022,NielsenChuang2010,Shor1996}.

Recently, a general framework called Gleipnir~\cite{Tao2021} was proposed to analyse and quantify the presence of quantum errors in quantum circuits. Gleipnir relies on MPS with truncated bond dimension to efficiently represent quantum states as well as compute a certain distance metric, more specifically a diamond norm, that is used to estimate quantum errors in quantum states. To compute such diamond norm, Gleipnir solves an associated semi-definite programming (SDP) problem whose size scales exponentially with the maximum number of qubits used by the quantum gates in a given circuit. But Gleipnir assumes all quantum gates have two input qubits at most as NISQ computers are unlikely to support more. Therefore, Gleipnir assumes such SDP is constant-sized so that the diamond norm can be computed efficiently. It also assumes that the noiseless quantum state (used as reference to compute said diamond norm) is known in advance. Under a simple bit-flip noise model, it is shown~\cite{Tao2021} that Gleipnir provides error bounds $15\%$ to $30\%$ tighter than previously known diamond norm estimates, as benchmarked on quantum approximate optimisation algorithms (i.e.~a form of VQA, see Section~\ref{opt:vqa-plateaus}) and a Lenz-Ising model with up to 100 qubits and $2{\;}265$ quantum gates. Furthermore, based on the proposed diamond norm, Gleipnir can guide quantum program compilers on how to best map physical qubits to logical qubits for noise reduction given a specific quantum hardware architecture. An example of this is shown for three and five qubit GHZ states on IBM's Boeblingen 20-qubit superconducting quantum computer.

The scalability of QEC methods has also been improved using higher {\linebreak} dimensional TNs like PEPS~\cite{darmawan2017tensor}. Using exact and approximate PEPS, the {\linebreak} authors simulate error correction via surface codes with more than 100 data qubits under two realistic noise models: amplitude-damping and systematic-rotation noise models. QEC methods based on other TN topologies including MERA have been explored by Ferris and Poulin~\cite{ferris2014tensor}.

\section{Discussion}\label{discussion}
TNs can speed up and reduce memory usage of classical simulations for {\linebreak} certain quantum circuits, while sacrificing accuracy by approximately rather than exactly representing quantum states. Their computational efficiency, the {\linebreak} expressiveness to represent general quantum physical systems, and scalability via massively parallel hardware, are well-known advantages of TN methods which make them a viable alternative to full state-vector representations. This is reflected in the wide range of TN applications developed particularly during the last decade, as reviewed in this paper.

In practice, however, whether a given TN shows performance advantages {\linebreak} depends on many different factors including: the specific TN structure, choice of TN contraction algorithm, critical TN parameters like bond dimension, {\linebreak} performance metrics benchmarked as well as properties of the quantum circuit itself being simulated like circuit depth and entanglement.

For example, compared with TTN and PEPS, image classification models {\linebreak} based on MPS require fewer training parameters, especially if implemented on quantum computers instead of classically simulating them (see Section~\ref{ml}). Yet novel models based on MPS, TTN or PEPS, struggle to outperform or even match state-of-the-art convolutional neural networks in terms of classification test accuracy. Nevertheless, there exist applications in quantum many-body physics where certain TNs outperform others generally, for example: TTN for finding ground states of certain tree-shaped molecules (Section~\ref{materials:treeshape}), and PEPS for finding ground states of transverse-field Ising models with spins arranged on a two-dimensional grid as recently shown in quantum advantage experiments (Section~\ref{qc:advantage}). By contrast, TN applications based on MERA are scarce across all research domains reviewed, arguably due to: the lack of efficient contraction algorithms for MERA and the fact that already many high-dimensional quantum physical systems can be represented via TTN or PEPS at a lower computational cost.

Not surprisingly, one of the main current challenges is designing standardised benchmark suites and good practices to rigorously evaluate the performance of TN applications. More so, given the vast number of quantum circuit simulators {\linebreak} available~\cite{Psarras2022,YoungSceseEbnenasir2023}. In fact, this challenge is not specific to TN software but common to quantum-computing software in general~\cite{FingerhuthBabejWittek2018}. Some pitfalls in {\linebreak} experiment settings we found are, for example: benchmarking only one aspect of the application (e.g.~test accuracy for applications in image classification); using unreliable metrics for classification like AUROC~\cite{ChiccoJurman2023}; and, measuring wall-clock time but not number of cost/energy function evaluations, which is a more robust hardware-agnostic metric and often used in runtime algorithm analysis.{\linebreak} We expect that future TN applications will benefit from recent developments in benchmark suites for quantum computing applications and related software~\cite{Dumitrescu2018,Finzgar2022,Jamadagni2024,Lubinski2023}.

The rapid growth of TN applications and related software during the last decade has been enabled by the wealth of TN algorithms in the literature~\cite{Bañuls2023,Bridgeman2017,Evenbly2022,GrasedyckKressnerTobler2013,KoldaBader2009,OkunishiNishinoUeda2022,Orus2014,Orus2019,Schollwock2011}. However, one notably less explored yet promising direction for future research are applications of hybrid methods based on TNs and other known approaches to quantum-circuit simulation. Two potential candidates that we found are tensor-based decision diagrams \cite{Hong2022} and tensor-based circuit cutting techniques \cite{Guala2023}.

Overall, we believe this review provides a representative and up-to-date {\linebreak} account of state-of-the-art TN applications across many research domains. A summary of these can be found in the Tables~\ref{table:2}--\ref{table:4} below, where applications are separated by rows and the following column fields: the original bibliographic reference introducing the application; the research domain where the application {\linebreak} focuses as presented by the authors; a brief description of the application; and, the TN class (or classes) used in such application.

\begin{table}[H]
    \centering
    \small
    \begin{threeparttable}
    \begin{tabular}{@{}ccp{0.35\linewidth}c@{}}
        \toprule
        Research Domain & Tensor Network\tnote{1,2} & \multicolumn{1}{c}{Application Description} & Reference
        \\ \toprule
        \multirow[c]{4}{*}{Machine learning}
        & \multirow[c]{4}{*}{MPS}
        & One-class linear classifier for anomaly detection in MNIST and Fashion-MNIST greyscale images and tabular data
        & \multirow[c]{4}{*}{\cite{Wang2020}}
        \\ \midrule
        \multirow[c]{3}{*}{Machine learning}
        & \multirow[c]{3}{*}{TTN}
        & Binary classifier for MNIST greyscale and CIFAR-10 coloured images
        & \multirow[c]{3}{*}{\cite{LiuDing2019}}
        \\ \midrule
        \multirow[c]{3}{*}{Machine learning}
        & \multirow[c]{3}{*}{MPS, TTN, PEPS}
        & Multi-class classifier for MNIST and Fashion-MNIST greyscale images
        & \multirow[c]{3}{*}{\cite{ChenHao2023}}
        \\ \midrule
        \multirow[c]{3}{*}{Machine learning}
        & \multirow[c]{3}{*}{MPS}
        & Classification of quarks in calorimeter images generated at the Large Hadron Collider
        & \multirow[c]{3}{*}{\cite{ArazSpannowsky2021}}
        \\ \midrule
        \multirow[c]{3}{*}{Machine learning}
        & \multirow[c]{3}{*}{MPS, TTN, MERA}
        & Classification of quarks in calorimeter images generated at the Large Hadron Collider
        & \multirow[c]{3}{*}{\cite{ArazSpannowsky2022}}
        \\ \midrule
        \multirow[c]{3}{*}{Machine learning}
        & \multirow[c]{3}{*}{MPS, PEPS}
        & Multi-class classifier for Fashion-MNIST and COVID-19 X-ray chest  images
        & \multirow[c]{3}{*}{\cite{LiLai2023}}
        \\ \midrule
        \multirow[c]{2}{*}{Machine learning}
        & \multirow[c]{2}{*}{MPS, TTN}
        & Image generation of MNIST handwritten digits
        & \multirow[c]{2}{*}{\cite{Cheng2019}}
        \\ \midrule
        \multirow[c]{4}{*}{Machine learning}
        & \multirow[c]{4}{*}{PEPS}
        & Generation of phase diagram images for a two-dimensional, frustrated, bilayer, Heisenberg Hamiltonian model
        & \multirow[c]{4}{*}{\cite{Kottmann2021}}
        \\ \midrule
        \multirow[c]{2}{*}{Optimisation}
        & \multirow[c]{2}{*}{MPS}
        & Multi-period mean-variance portfolio optimisation
        & \multirow[c]{2}{*}{\cite{Mugel2022}}
        \\ \midrule
        \multirow[c]{3}{*}{Optimisation}
        & \multirow[c]{3}{*}{TTN}
        & Dose optimisation in intensity-modulated radiation {\linebreak}therapy for cancer treatment
        & \multirow[c]{3}{*}{\cite{Cavinato2021}}
        \\ \midrule
        \multirow[c]{3}{*}{Optimisation}
        & \multirow[c]{3}{*}{MPS}
        & Analysis of a variational classical-quantum algorithm for solving QUBO problems
        & \multirow[c]{3}{*}{\cite{Perelshtein2023}}
        \\ \midrule
        \multirow[c]{3}{*}{Optimisation}
        & \multirow[c]{3}{*}{MPS, TTN, MERA}
        & Analysis of barren plateaus in cost functions in variational quantum optimisation
        & \multirow[c]{3}{*}{\cite{MiaoBarthel2024}}
        \\ \bottomrule
    \end{tabular}
    \begin{tablenotes}
        \item [1] MPS: matrix product state; TTN: tree tensor network; PEPS: projected entangled-pair state; isoTNS: isometric tensor network state; MERA: multi-scale entanglement renormalisation ansatz.
        \item [2] This column includes any tensor network used in the referenced paper not just those proposed by the authors.
    \end{tablenotes}
    \caption{Tensor network applications.\label{table:2}}
    \end{threeparttable}
\end{table}
\begin{table}[H]
    \centering
    \small
    \begin{tabular}{@{} c >{\centering}p{0.20\linewidth} p{0.35\linewidth} c @{}}
        \toprule
        Research Domain & Tensor Network & \multicolumn{1}{c}{Application Description} & Reference
        \\ \toprule
        \multirow[c]{3}{*}{Materials science}
        & \multirow[c]{3}{*}{MPS}
        & Energy function minimisation for a Hubbard Hamiltonian model of artificial graphene
        & \multirow[c]{3}{*}{\cite{Obiol2022}}
        \\ \midrule
        \multirow[c]{4}{*}{Materials science}
        & \multirow[c]{4}{*}{PEPS}
        & Analysis of energy ground states for strontium copper borate, described by the Shastry-Sutherland model
        & \multirow[c]{4}{*}{\cite{CorbozMila2014,Shi2022}}
        \\ \midrule
        \multirow[c]{3}{*}{Materials science}
        & \multirow[c]{2}{*}{MPS, } \multirow[c]{2}{*}{2D isoTNS}
        & Computing thermal states for a two-dimensional transverse-field Ising Hamiltonian model
        & \multirow[c]{3}{*}{\cite{KadowPollmanKnap2023}}
        \\ \midrule
        \multirow[c]{4}{*}{Quantum chemistry}
        & \multirow[c]{4}{*}{MPS}
        & Analysis of hydrogen chains, torsional barrier of ethane and protein-ligand interactions in SARS-CoV-2
        & \multirow[c]{4}{*}{\cite{shang2023towards}}
        \\ \midrule
        \multirow[c]{3}{*}{Quantum chemistry}
        & \multirow[c]{3}{*}{MPS, TTN}
        & Energy function minimisation for a Hubbard Hamiltonian model of lithium fluoride
        & \multirow[c]{3}{*}{\cite{Murg2015}}
        \\ \midrule
        \multirow[c]{3}{*}{Quantum chemistry}
        & \multirow[c]{3}{*}{MPS, TTN}
        & Energy function minimisation for a Hubbard model of tree-shaped molecules
        & \multirow[c]{3}{*}{\cite{NakataniChan2013}}
        \\ \midrule
        \multirow[c]{3}{*}{Quantum simulation}
        & \multirow[c]{3}{*}{MPS, TTN}
        & Reducing simulation time for certain quantum circuits with a tree-shaped layout
        & \multirow[c]{3}{*}{\cite{Seitz2023}}
        \\ \midrule
        \multirow[c]{3}{*}{Quantum simulation}
        & \multirow[c]{3}{*}{PEPS}
        & Approximate simulation of random quantum circuits {\linebreak} including Google's Sycamore
        & \multirow[c]{3}{*}{\cite{Liu2021}}
        \\ \midrule
        \multirow[c]{2}{*}{Quantum simulation}
        & \multirow[c]{2}{*}{PEPS}
        & Exact simulation of random quantum circuits
        & \multirow[c]{2}{*}{\cite{guo2019general}}
        \\ \midrule
        \multirow[c]{4}{*}{Quantum simulation}
        & \multirow[c]{3}{*}{MPS, } \multirow[c]{3}{*}{2D isoTNS}
        & Benchmark the quantum processor IBM Eagle for a two-dimensional transverse-field Ising Hamiltonian model
        & \multirow[c]{4}{*}{\cite{Kim2023}}
        \\ \bottomrule
    \end{tabular}
    \caption{Tensor network applications (continued).\label{table:3}}
\end{table}
\begin{table}[H]
    \centering
    \small
    \begin{threeparttable}
    \begin{tabular}{@{} c >{\centering}p{0.20\linewidth} p{0.35\linewidth} c @{}}
        \toprule
        Research Domain\tnote{1} & Tensor Network & \multicolumn{1}{c}{Application Description} & Reference
        \\ \toprule
        \multirow[c]{4}{*}{Quantum simulation}
        & \multirow[c]{3}{*}{MPS, PEPS,} \multirow[c]{3}{*}{2D isoTNS}
        & Benchmark the quantum processor IBM Eagle for a two-dimensional transverse-field Ising Hamiltonian model
        & \multirow[c]{4}{*}{\cite{Patra2024}}
        \\ \midrule
        \multirow[c]{3}{*}{CFD}
        & \multirow[c]{3}{*}{MPS}
        & Solving Navier-Stokes equations for wall-bounded flows in two spatial dimensions
        & \multirow[c]{3}{*}{\cite{KiffnerJaksch2023}}
        \\ \midrule
        \multirow[c]{3}{*}{QE correction}
        & \multirow[c]{3}{*}{PEPS}
        & Noise cancellation in quantum circuits made of non-Clifford gates via surface codes
        & \multirow[c]{3}{*}{\cite{darmawan2017tensor}}
        \\ \midrule
        \multirow[c]{2}{*}{QE analysis}
        & \multirow[c]{2}{*}{MPS}
        & Error bounds estimation in noisy quantum programs
        & \multirow[c]{2}{*}{\cite{Tao2021}}
        \\ \bottomrule
    \end{tabular}
    \begin{tablenotes}
        \item [1] CFD: computational fluid dynamics; QE: quantum error.
    \end{tablenotes}
    \caption{Tensor network applications (continued).\label{table:4}}
    \end{threeparttable}
\end{table}

\section*{Acknowledgements}
We are deeply grateful to Dr Artur Garcia Sáez, leading researcher at the Barcelona Supercomputing Center in Spain, for his helpful feedback and {\linebreak} expertise on tensor networks which have positively contributed to the draft of this paper. We also appreciate his suggestions of relevant papers and topics regarding tensor network applications in quantum computing.
\bibliographystyle{acm}
\bibliography{refs}
\addcontentsline{toc}{section}{\refname}
\end{document}